\documentclass[runningheads]{llncs}
\usepackage[colorlinks,linkcolor=blue,anchorcolor=blue,citecolor=blue]{hyperref}
\usepackage{graphicx}
\usepackage{amssymb}
\usepackage{amsmath}
\usepackage{array}
\usepackage{cite}
\usepackage{hyperref}
\usepackage[misc,geometry]{ifsym}

\newcolumntype{N}{>{\centering\arraybackslash}m{.5in}}
\newcolumntype{G}{>{\centering\arraybackslash}m{2in}}

\def\A{\mathbf{A}}

\def\S{\mathbf{S}}
\def\X{\mathbf{X}}
\def\Y{\mathbf{Y}}
\begin{document}
\title{Symmetry-Enhanced Attention Network for Acute Ischemic Infarct Segmentation with Non–Contrast CT Images}
\titlerunning{Symmetry Enhanced Attention Network}
%
\author{
Kongming Liang\inst{1} \and
Kai Han\inst{2} \and
Xiuli Li\inst{2} \and
Xiaoqing Cheng\inst{3} \and
Yiming Li\inst{2} \and \\
Yizhou Wang\inst{4} \and
Yizhou Yu\inst{2,5} $^{\left(\textrm{\Letter}\right)}$
}

%
\authorrunning{K. Liang et al.}
%
\institute{
$^1$Pattern Recognition and Intelligent System Laboratory, School of Artificial Intelligence, Beijing University of Posts and Telecommunications, Beijing, China\\
$^2$Deepwise AI Lab, Beijing, China\\
$^3$Department of Medical Imaging, Jinling Hospital, Nanjing University School of Medicine, Nanjing, Jiangsu, China\\
$^4$Department of  Computer Science and Technology, Peking University, Beijing, China\\
$^5$The University of Hong Kong, Pokfulam, Hong Kong \\
\email{yizhouy@acm.org}}
%
\maketitle              
\begin{abstract}
Quantitative estimation of the acute ischemic infarct is crucial to improve neurological outcomes of the patients with stroke symptoms. Since the density of lesions is subtle and can be confounded by normal physiologic changes, anatomical asymmetry provides useful information to differentiate the ischemic and healthy brain tissue. In this paper, we propose a symmetry enhanced attention network (SEAN) for acute ischemic infarct segmentation. Our proposed network automatically transforms an input CT image into the standard space where the brain tissue is bilaterally symmetric. The transformed image is further processed by a U-shape network integrated with the proposed symmetry enhanced attention for pixel-wise labelling. The symmetry enhanced attention can efficiently capture context information from the opposite side of the image by estimating long-range dependencies. Experimental results show that the proposed SEAN outperforms some symmetry-based state-of-the-art methods in terms of both dice coefficient and infarct localization.

\end{abstract}
\section{Introduction}
\begin{figure}[t]
\centering
\includegraphics[scale=0.6]{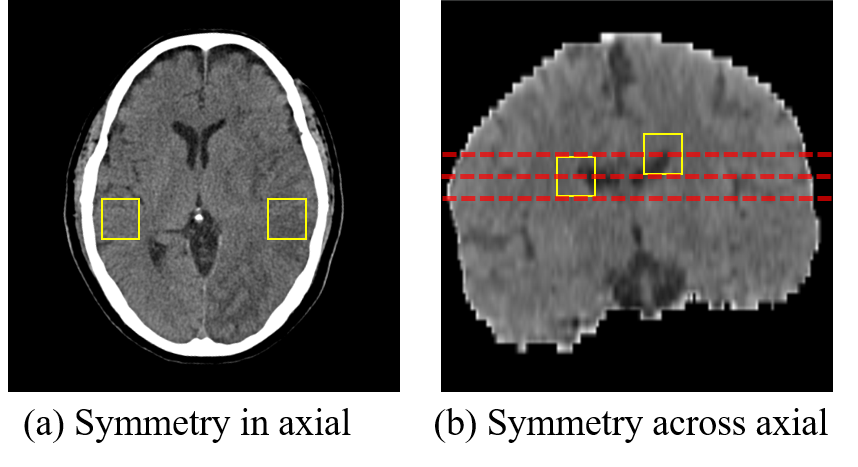}
\vskip -0.1in
\caption{The anatomical symmetry of the brain CT images. (a) and (b) show the two symmetrical patches of the image in axial view and across axial view. The dotted red line in (b) denotes the slice from axial view. Due to the rotation of patient's head, the symmetrical landmarks may appear on different images in the axial view.} \label{fig_SYM}
\vskip -0.2in
\end{figure}

Acute ischemic stroke is one of the leading causes of death and disability worldwide and imposes an enormous burden for the health care system \cite{katan2018global}. The use of pretreatment neuroimaging is critical to improve neurological outcomes of patients with stroke symptoms. Compared to MRI, non-contrast head CT scan is commonly used as the initial imaging because of its wide availability and low acquisition time. To interpret early infarct signs in CT, the Alberta Stroke Program Early CT Score (ASPECTS) evaluation was proposed at the beginning in the 2000s~\cite{2000-Lancet-Philip} and has found increasing acceptance in clinical practice. However, ASPECTS evaluation is only an approximation of the assessment of early ischemic changes. Since the density of lesions is subtle and can be confounded by normal physiologic changes, quantitative estimation of acute ischemic infarct is challenging. In clinical practice \cite{2015-Khan_Academy}, bilaterally symmetric (illustrated in Fig.~\ref{fig_SYM}) provides useful information for the identification of acute ischemic infarct.

Anatomical asymmetry has been utilized in previous works to localize and segment the abnormal regions for neuroimaging analysis.  \cite{2020-Radiology-Wu,2016-ICIP-Wang} leverage the symmetry by adding extra information beyond the input image. \cite{2020-Radiology-Wu} calculates the differences of each voxel by subtracting the original brain from the mirrored brain. The difference map is further used as the input to train a random forest classifier to yield lesion segmentation. \cite{2016-ICIP-Wang} extracts both the original patch and its symmetric patch, and feeds them into the network simultaneously. Except for calculating the asymmetry on image-level, \cite{2019-ISBI-Arko,2019-MRI-Liu,2017-NIPSW-Zhang} propose to explore feature-level fusion of the two symmetry regions. For instance, two-branch networks (e.g. siamese network) can learn the features of left and right hemispheres and measure the difference between the features of two hemispheres to analyze abnormalities such as Alzheimer's disease~\cite{2019-MRI-Liu}, ischemic stroke \cite{2019-ISBI-Arko,2019-MICCAI-kuang} and brain tumors \cite{2017-NIPSW-Zhang}. Even though the pixel-wise difference is widely used in previous methods, it can not efficiently exploit the bilaterally symmetric information due to the limitation of context modeling. In addition, all the above methods need the input images to be already calibrated which cannot be guaranteed in practice.

In this paper, a symmetry enhanced attention network (SEAN) is proposed for acute ischemic infarct segmentation. The proposed SEAN can automatically transform an input image into the standard space without any human supervision. The transformed image is further processed by a U-shape network that contains encoding and decoding stages. Different from the original design ~\cite{2015-MICCAI-ronneberger}, the encoder performs 3d convolution to leverage context information of adjacent images in axial. Then, a symmetry enhanced attention module is integrated between the encoding and decoding stages to efficiently model the anatomical symmetry. In summary, the main contributions of our paper are as follows.

\begin{enumerate}
	\item A symmetry enhanced attention is proposed to capture both in-axial and cross-axial symmetry information by explicitly estimating the long-range dependencies.
	\item A symmetry-based alignment network is proposed to transform an input image as bilaterally symmetric in axial without any human supervision.
	\item We release the dataset at \url{https://github.com/GriffinLiang/AISD}.
\end{enumerate}

\section{Method}
\begin{figure}[t]
\centering
\includegraphics[scale=0.45]{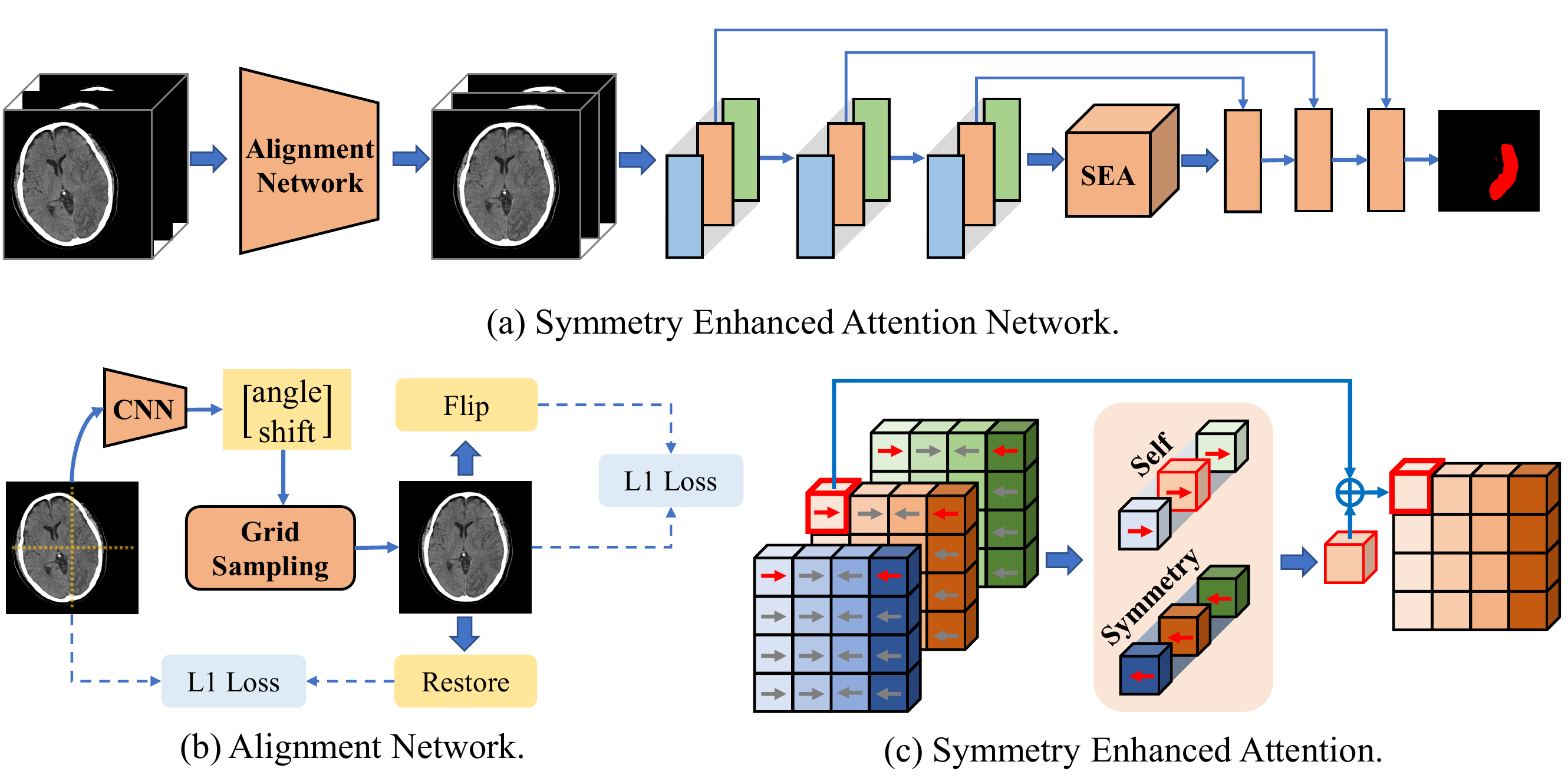}
\caption{Overview of our proposed network architecture.} \label{fig_SEAN}
\end{figure}

In this section, we first introduce the symmetry based alignment network in Sec.~\ref{sec:align} and give the detail information of how to make the input image bilaterally symmetric in axial. Then we define the structure of the proposed symmetry enhanced attention network which can capture both the in-planar and across-planar symmetry information for ischemic infarct segmentation in Sec.~\ref{sec:network}. The whole pipeline is shown in Fig.~\ref{fig_SEAN}.

\subsection{Symmetry based Alignment}
\label{sec:align}
Since the poses of patients are arbitrary when they perform CT scans, the brain images are usually not in standard space. In order to effectively use the symmetry of the brain, we attempt to align the image to keep the region of brain in the center of the image and horizontally symmetrical. However, traditional registration based method can not be applied in clinical practice due to the high time complexity. Therefore, we proposed a Symmetry based Alignment Network as show in Fig.~\ref{fig_SEAN} which can automatically align the brain images with only the information of images itself. Inspired by Spatial Transformer Networks \cite{2015-NIPS-jaderberg}, we design the symmetry based alignment network as: c2d[32,7,7]-relu-max2d[2,2]-c2d[32,5,5]-relu-max2d[2,2]-fc[3] where c2d[$n$,$c_w$,$c_h$] denotes a 2d convolutional layer with n filters of size $c_w \times c_h$, max2d[$s_h,s_w$] is a 2d max-pooling layer with the kernel size and the stride as $s_h \times s_w$, fc[$n$] is a fully connected layer with $n$ units. The output of the network is interpreted as the parameters $\alpha$ (rotation, horizontal shift and vertical shift) of rigid transformation matrix.

Given an input volume, we define $\A_i$ as the $i$-th slice in the axial view. During training, the output parameters $\alpha$ is applied to the input slice $\A_i^t = f_{\alpha}(\A_i)$ using parameterised sampling grid. Then we generate the flipped version of $\A_i^t$ as $\Tilde{\A}_i^t$. The total loss is designed as the following:

\begin{equation}
\label{eqn:align}
\begin{aligned}
L_{a} = 
L_{1}(\Tilde{\A}_i^t, \A_i^t) +
L_{1}(f_{\alpha}^{-1}(\A_i^t), \A_i),
\end{aligned}
\end{equation}
where $f_{\alpha}^{-1}(\cdot)$ denotes the inverse transformation function and $L_1$ denotes the L1 loss. We define the first term of Eqn.~(\ref{eqn:align}) as symmetry loss which is the L1 distance between the aligned image and its horizontally flipped image. The symmetry loss is based on the assumption that the difference between the image of brain and its horizontally flipped version will be minimized when the brain is perfectly aligned. In addition, we need to add constraints to the symmetry loss to avoid trivial solution where the alignment network can simply transform the brain region out of the input image. Therefore, a restoration loss is defined as the L1 distance between the restored image and the original image to learn useful transformation parameters.

\subsection{Symmetry based Segmentation}
\label{sec:network}
The proposed segmentation network adopts the structure of UNet~\cite{2015-MICCAI-ronneberger} which is mainly composed of two parts: the encoder stage and the decoder stage. Inspired by \cite{2019-MICCAI-fang}, we use 3D convolutions as the basic encoding block to keep the context information from adjacent images in axial view. For the decoding stage, the middle plane of input volume is retained as the target image and upsampled to the original resolution for pixel-wise labelling. We name the above network as HybridUnet. Finally, we cascade the last encoding block with the symmetry attention module. In this way, the feature representation can be enhanced by its symmetry information to efficiently assess the presence and extent of ischemic infarct.

To exploit the context information of the $i$-th axial image, HybridUnet takes its adjacent images $\{\A_{i+t}| t=-T, \cdots, T\}$ as the input. The input images are firstly processed by the 3d encoder. We design the encoder block as: c3d-bn-relu-c3d-bn-relu-max3d where c3d denotes 3d convolutional layer, bn denotes 3d batchnorm layer and max denotes a 3d max-pooling layer. The encoder stage contains five encoder blocks. The output feature from the last encoder block is denoted as $\X_{i} \in \mathbb{R}^{C \times H \times W}$ for the input image $\A_i$ where $H$ and $W$ represent the height and width of the output feature respectively and $C$ is the number of the output channels. The output feature $\X_{i}$ is further processed by symmetry attention module in Sec.~\ref{sec:sea} and further processed by four decoding blocks with the same structure as the original UNet. We train the proposed SEAN by minimizing the combination of the generalized Dice loss and the cross-entropy loss in an end-to-end manner.

\subsubsection{Symmetry Enhanced Attention.}
\label{sec:sea}
The symmetry information is hard to explore by the conventional operation which only processes a local neighborhood in space. Since the input position and its symmetrical position usually have long spatial interval distances, local operations need to be applied repeatedly to capture such long-range dependencies. As mentioned in \cite{2015-CVPR-He}, stacking multiple convolution operations is computationally inefficient and increases the difficulty of optimization. To compensate for the drawback of convolution operation, we propose to model the relationships between symmetrical position with attention mechanism~\cite{2018-CVPR-wang}. Given an input feature map $\X_i \in \mathbb{R}^{C \times H \times W}$, we first divide it into $P \times Q$ partitions as below,

\begin{equation}
\label{eqn:x}
\begin{aligned}
\X_i = \begin{bmatrix}
 \X_{i,1,1} & \X_{i,1,2} & \cdots & \X_{i,1,Q} \\ 
 \X_{i,2,1} & \X_{i,2,2} & \cdots & \X_{i,2,Q} \\ 
 \vdots & \vdots & \ddots & \vdots \\ 
 \X_{i,P,1} & \X_{i,P,2} & \cdots & \X_{i,P,Q}
\end{bmatrix},
\end{aligned}
\end{equation}
where $\X_{i,j,k} \in \mathbb{R}^{C \times H^{'} \times W^{'}}$ is a subset of $\X_i$ ($H=H^{'} \times P$ and $W=W^{'} \times Q$). Then we flip $\X_i$ horizontally to generate its mirrored feature map $\Tilde{\X}_i$. Therefore, the symmetrical partition of $\X_{i,j,k}$ can be denoted as $\Tilde{\X}_{i,j,k}$.

For the input partition $\X_{i,j,k}$, its symmetry enhanced attention is composed of the self-attention module and the symmetry-attention module:

\begin{equation}
\label{eqn:self_att}
\begin{aligned}
\S_{i,j,k}^t & = Softmax(\frac{\theta(\X_{i,j,k})^{\top}\phi(\X_{i+t,j,k})}{\sqrt{d}}),\\
\Tilde{\S}_{i,j,k}^t & = Softmax(\frac{\theta(\X_{i,j,k})^{\top}\phi(\Tilde{\X}_{i+t,j,k})}{\sqrt{d}}),\\
\Y_{i,j,k} & = 
\left ( \sum_{t=-T}^T \S_{i,j,k}^t \cdot g(\X_{i+t,j,k})^{\top}
+ \sum_{t=-T}^T \Tilde{\S}_{i,j,k}^t \cdot h(\Tilde{\X}_{i+t,j,k})^{\top}
\right )^{\top},
\end{aligned}
\end{equation}
where $\S_{i,j,k}^t, \Tilde{\S}_{i,j,k}^t \in \mathbb{R}^{N^{'} \times N^{'}} (N^{'}=H^{'} \times W^{'})$ are the similarity matrix of the self attention and symmetry attention respectively. $\theta(\cdot)$ and $\phi(\cdot)$ perform convolution operations to reduce the number of input channels to $d$ (e.g. $d=\frac{C}{2}$) and reshape the output to $\mathbb{R}^{d \times H^{'}W^{'}}$. We use $\sqrt{d}$ as a scaling factor for the inner product to solve the small gradient problem of softmax function. $\X_{i+t,j,k}$ and $\Tilde{\X}_{i+t,j,k}$ are also fed into $g(\cdot)$ and $h(\cdot)$ to compute the new representation by convolution operations and reshape the output feature map to $\mathbb{R}^{\frac{C}{2} \times H^{'}W^{'}}$. The symmetry enhance feature $\Y_{i,j,k} \in \mathbb{R}^{C \times H^{'}W^{'}} $ is reshaped to $C \times H^{'} \times W^{'}$ and further considered as the residual mapping of $\X_{i,j,k}$ to acquire the final output of the symmetry enhanced attention.

\section{Experiments}
In this section, we first introduce the data acquisition and evaluation indicators of our model in Sec.~\ref{sec:exp_setup}. Then we show the detail information of implementation in Sec.~\ref{sec:exp_implementation}. Finally, we compare our approach with the state-of-the-art methods and conduct extensive ablation studies in Sec.~\ref{sec:exp_results}.

\subsection{Experiment setup}
\label{sec:exp_setup}
\subsubsection{Data Acquisition.} We obtain 397 Non–Contrast-enhanced CT (NCCT) scans of acute ischemic stroke with the interval from symptom onset to CT less than 24 hours. The patients underwent diffusion-weighted MRI (DWI) within 24 hours after taking the CT. The slice thickness of NCCT is 5mm. We name the above CT-MRI pairs as acute ischemic stroke dataset (AISD). 345 scans are used to train and validate the model, and the remaining 52 scans are used for testing. Ischemic lesions are manually contoured on NCCT by a doctor using MRI scans as the reference standard. Then a senior doctor double-reviews the labels.

\subsubsection{Evaluation Metrics.} To quantitatively evaluate the result of our proposed symmetry based alignment network, we compare the output transformation parameters with the human annotated rotation and offset. Specifically, a doctor annotates the beginning and end point of cerebral falx on the middle slice of each CT scan. The offset angle and the center of brain region are further calculated as ground truth. We use the average difference of the rotation angle and the offset distance in the horizontal direction between the model's output and the ground truth for each data. As for the segmentation results, we utilize Dice coefficient to quantitatively evaluate the performance. 
In addition, we also calculate infarct-level evaluation metrics such as recall, precision and F1 score. To evaluate the clinical value, Pearson correlation between the estimated ASPECTS and the ground-truth is also performed.

\subsubsection{Comparison Methods.} We compare the proposed SEAN with following methods: 1) No symmetry modelling method (Unet) which is a Vanilla Unet without considering the symmetry information; 2) Symmetry modelling on image-level (Unet-IM-L1) \cite{2020-Radiology-Wu} which calculates the bilateral density $L_1$ difference between symmetric brain regions as one of the input; 3) Symmetry modelling on feature-level by concatenation (Unet-FT-CC) which takes the original image and its flipped image as the input of Unet and concatenates the two output features from the last encoder; 4) Symmetry on feature-level with L1 distance (Unet-FT-L1) \cite{2017-NIPSW-Zhang} which takes the distance map between the feature from the last encoder and its flipped version as an extra information. We also implement the above methods using HybridUnet \cite{2019-MICCAI-fang} as the backbone for further comparison.

\subsection{Implementation Details}
\label{sec:exp_implementation}
Our implementation is based on Pytorch framework. For data pre-processing, we truncate the raw intensity values to the range [40, 100] HU and normalize each raw CT case to have zero mean and unit variance. The Adam optimizer is used to train the model with parameters $\beta_{1}=0.9, \beta_{2}=0.99$ for 150 epochs. And we set the base learning rate as $1 \times 10^{-4}$ and deploy a poly learning rate policy where the initial learning rate is multiplied by $(1-\frac{iter}{total\_iter})^{power}$ and $power=0.9$ after each iteration.

\subsection{Results and Discussions}
\label{sec:exp_results}
\subsubsection{Efficacy of Alignment Network.} We evaluate the similarity of the rotation angle and the offset distance in the horizontal direction between the model's output and the ground truth on test data (52 patients). The error of rotation angle is 3 degrees and the error of shift distance is 5 pixels. From the results, we can see that the output of the symmetry based alignment network is very close to the ground truth even though the proposed method is fully unsupervised.
Besides, we register the NCCT images to the standard brain template to align the images as described in \cite{2020-Radiology-Wu}. This operation cost 134 seconds per patient on average, while the time consumption of our proposed symmetry based alignment network is only 0.46 seconds per patient on average. The registration method is based on Advanced Normalization Tools(ANTS)~\cite{avants2009advanced}, and all experiments are performed on a Linux server with Intel(R) Xeon(R) CPU E5-2697 v4 @ 2.30GHz and a NVIDIA 2080ti GPU.

\begin{table}[t]  
\centering  
\caption{Quantitative results on the AIS dataset.}  
\label{tab_res}
 \begin{tabular}{l|c|c||c|c|c|c}  
     \hline
     \hline
        Network & Fusion-Level & Fusion-Type & Dice & F1 & Recall & Precision\\
       \hline
        Unet & N/A & N/A & 0.4588 & 0.5105 & 0.5019 & 0.5196 \\
        Unet-IM-L1 & Image & L1 & 0.5035 & 0.5457 & 0.5318 & 0.5603 \\
        Unet-FT-L1 & Feature & L1 & 0.5121 & 0.567 & 0.5468 & 0.5888 \\
        Unet-FT-CC & Feature & Concat & 0.5354 & 0.572 & 0.5655 & 0.5786 \\
       \hline
        HybridUnet & N/A & N/A & 0.4952 & 0.5433 & 0.6105 & 0.4895 \\
        HybridUnet-IM-L1 & Image & L1 & 0.5437 & 0.5992 & 0.5581 & 0.6471 \\
        HybridUnet-FT-L1 & Feature & L1 & 0.5445 & 0.5982 & 0.5543 & 0.6497 \\
        HybridUnet-FT-CC & Feature & Concat & 0.5577 & 0.6015 & 0.5431 & 0.6742 \\
     \hline
       SEAN & Feature & Attention & 0.5784 & 0.6218 & 0.5880 & 0.6597 \\
       \hline
       \hline
   \end{tabular}
\vskip -0.2in
\end{table}

\subsubsection{Efficacy of SEAN.}
\begin{figure}[t]
\centering
\includegraphics[scale=0.52]{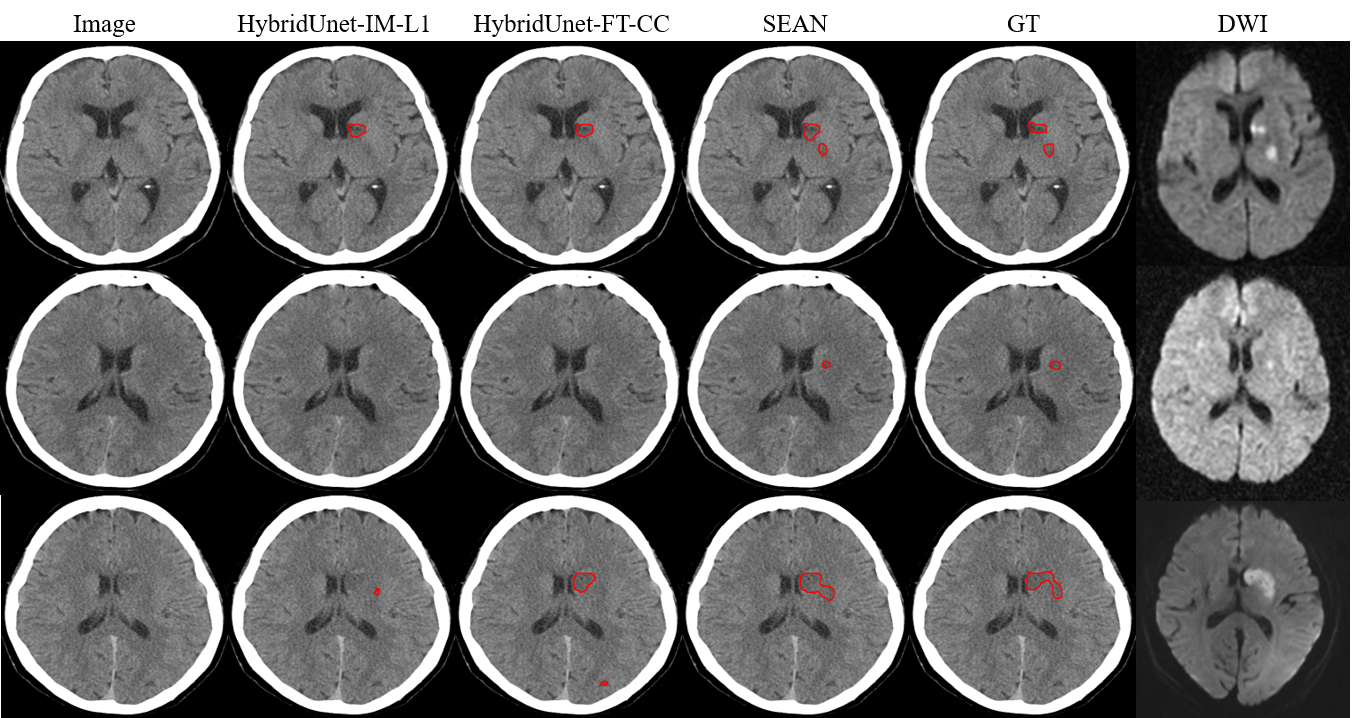}
\caption{Qualitative comparison of different methods.} \label{fig_vis}
\end{figure}

As shown in Tab.~\ref{tab_res}, the proposed SEAN achieves the highest performance (Dice: 0.5784, F1: 0.6218) compared to the other methods. Since the proposed SEAN benefits from leveraging both the in-planar and across-planar symmetry information, it can differentiate the infarct and the normal physiologic change more efficiently. In general, symmetry based  methods observed significant improvements according to both Dice and F1. This is also consistent with doctors' habit in clinical practice. For the effectiveness of the backbone network, HybridUnet achieves better performance than the original Unet, which demonstrates the importance of context information from adjacent images. Feature-level method outperforms image-level method, since the feature-level method is robust to the misalignment of the input image. We conduct ablation studies of SEAN and show the results in Tab.~\ref{tab_ablation}. The influence of the proposed alignment network and the two type of attention mechanism: self-attention and symmetry-attention is investigated. According to the results, it can be seen that the proposed alignment network improves over the baseline for a large margin. We can also see that symmetry enhanced attention yields a higher increase in both Dice and F1 comparing to only using self-attention. We also show some qualitative examples in Fig.~\ref{fig_vis}.


\begin{table}[t]  
\centering  
\caption{Ablation study of SEAN on AISD.}  
\label{tab_ablation}
 \begin{tabular}{l|c|c|c|c}  
     \hline
     \hline
        Method & Dice & F1 & Recall & Precision \\
      \hline
        baseline & 0.4952 & 0.5433 & 0.6105 & 0.4895 \\
      \hline  
        + Ours (Align)  & 0.5281 & 0.5767 & 0.5506 & 0.6056 \\
        + Ours (Align+Self)  & 0.5635 & 0.5834 & 0.5506 & 0.6205 \\
        + Ours (Align+Self+Sym)  & 0.5784 & 0.6218 & 0.5880 & 0.6597 \\
      \hline
      \hline
  \end{tabular}
\end{table}

\subsubsection{Efficacy of Clinical Usage.} To validate the clinical efficacy of the proposed SEAN, Pearson correlation between the estimated ASPECTS and the ground-truth is performed. A standard template with ASPECTS regions in the Montreal Neurologic Institute space \cite{evans19933d} is registered to all NCCT images by performing affine transformation using ANTS. The Pearson correlation between the scores estimated by SEAN and doctor is 0.75, which further indicates the efficiency of the proposed method.

\section{Conclusion}
In this paper, we propose a symmetry enhanced attention network (SEAN) for acute ischemic infarct segmentation. The proposed network calibrates an input CT image and capture bilateral symmetry information by explicitly estimating the long-range dependencies. With the seamless integration, the proposed symmetry enhanced attention can be applied to any lesion segmentation task. Experimental results on acute ischemic stroke dataset (AISD) show that the proposed SEAN outperforms some symmetry-based state-of-the-art methods in terms of both dice coefficient and infarct localization. The acute ischemic stroke dataset (AISD) is published for future study.

\subsubsection{Acknowledgments}
This work was supported in part by following grants, MOST-2018AAA0102004, NSFC-62061136001, the Key Program of Beijing Municipal Natural Science Foundation(7191003), and the Key Projects of the National Natural Science Foundation of China (81830057).

%
%
%

\bibliographystyle{splncs04}
\bibliography{mybibliography}
\end{document}


\begin{figure}[t]
\centering
\includegraphics[scale=0.55]{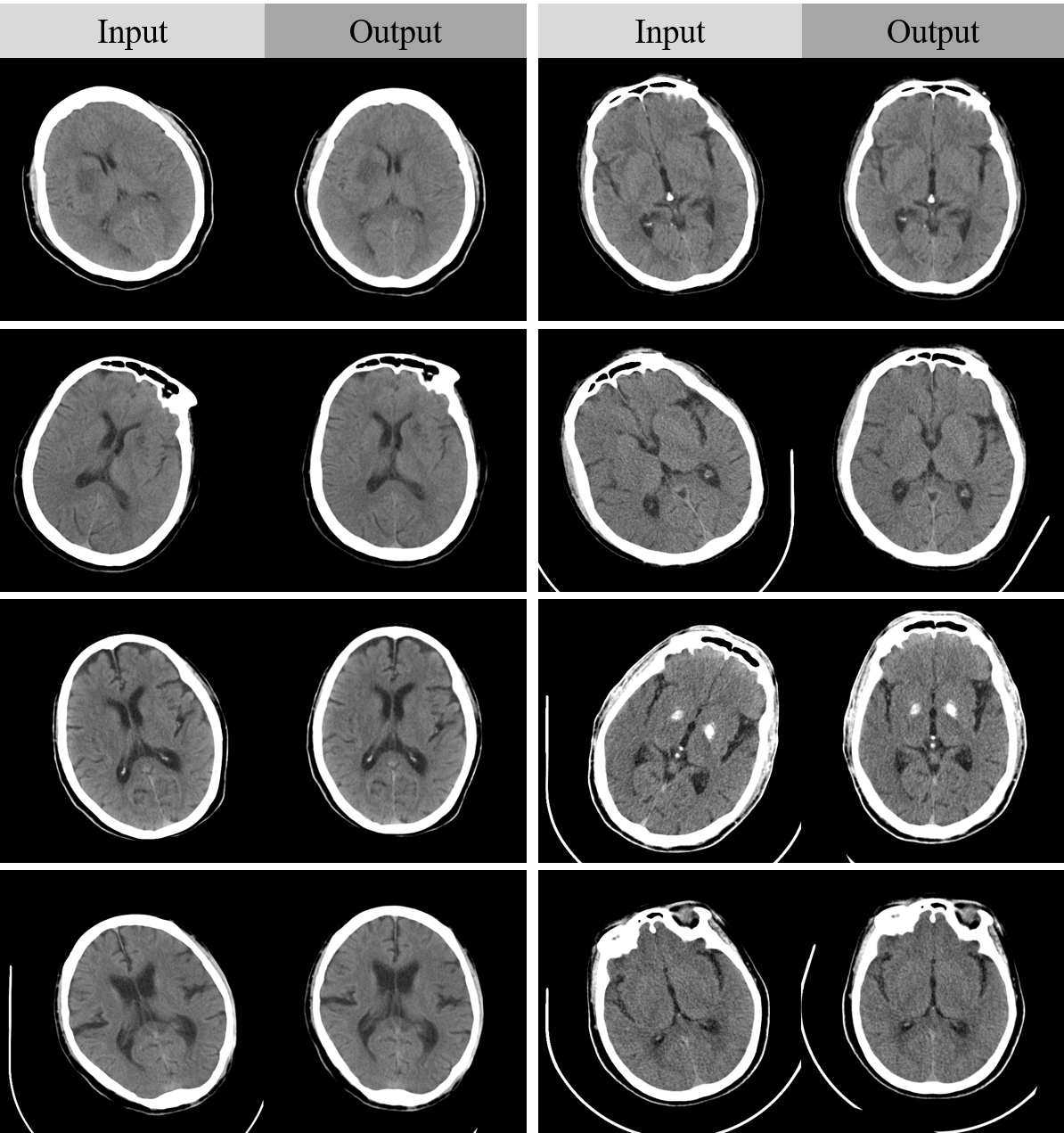}
\caption{Qualitative results of the proposed symmetry alignment network.}
\end{figure}

\begin{figure}[t]
\centering
\includegraphics[scale=0.4]{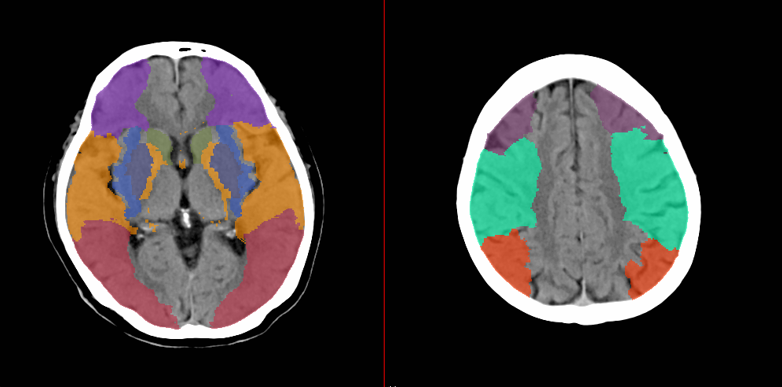}
\caption{ASPECTS regions are consist of 10 different parts which are symmetrical with respect to the cerebral falx.}
\end{figure}